\documentclass[aps,prd,twocolumn,superscriptaddress,floatfix]{revtex4-2}
\usepackage{graphicx}
\usepackage{amsmath}
\usepackage{amsfonts}
\usepackage{color}
\usepackage{bm}
\usepackage{bbm}
\usepackage{txfonts}
\usepackage{array}
\usepackage{amssymb}
\usepackage{ragged2e}
\usepackage{hyperref}
\usepackage{xcolor}

\hypersetup{
    colorlinks=true,
    linkcolor=green,
    filecolor=green,
    urlcolor=blue,
    citecolor=blue,
}

\renewcommand{\raggedright}{\leftskip=0pt \rightskip=0pt plus 0cm}

\begin{document}

\title{Schwinger Pair Production in Superposed Electric-Field Pulses with Distinct Spatial Widths}
\author{Abduweli Abdukerim}
\email{abduwelee@snnu.edu.cn}
\email{aasci@ksu.edu.cn}
\affiliation{College of Physics and Electrical Engineering, Kashi University, Kashgar 844000, China}

%\date{\today}

\begin{abstract}
Electron--positron pair production in spatially and temporally inhomogeneous electric fields is investigated in $1+1$ dimensions within the Dirac--Heisenberg--Wigner (DHW) formalism. We consider a composite field formed by two Gaussian pulses with independently adjustable spatial widths, $\lambda_1$ and $\lambda_2$, extending commonly used combined-pulse configurations in which the constituent pulses share the same spatial envelope.
Both temporally overlapping and temporally separated configurations are analyzed. For overlapping pulses, decreasing $\lambda_2$ at fixed $\lambda_1$ enlarges the weak-field region and enhances the low-momentum part of the particle spectrum. Oscillatory temporal profiles further increase the particle yield while reducing the net momentum transferred to the particles. For temporally separated pulses, the spatial width of the delayed pulse determines the extent to which it accelerates particles created by the earlier pulse. A narrow delayed pulse affects only a limited fraction of these particles, whereas a broader pulse substantially reshapes the final momentum distribution. These results identify the spatial-width disparity between the constituent pulses as an additional control parameter for Schwinger pair production in multicomponent electric fields.
\end{abstract}
\maketitle

\section{Introduction}

Electron--positron pair production from the vacuum in a strong electric field is a paradigmatic nonperturbative prediction of quantum electrodynamics (QED). Its theoretical foundations were established by the early work of Dirac, Klein, Sauter, Heisenberg and Euler, and Schwinger \cite{dirac,klein,Sauter,Heisenberg,Schwinger1951}. Direct observation of nonperturbative vacuum decay remains challenging because the characteristic critical field exceeds the strength of presently available quasistatic fields. Nevertheless, early proposals explored the possibility of vacuum pair production using x-ray free-electron lasers \cite{RINGWALD2001107,Alkofer2001}, while advances in high-intensity laser facilities have broadened the experimental prospects for strong-field QED \cite{10.1063/1.4909613,Fedotov2023Review}. In parallel, laser-driven experiments have demonstrated the generation of dense electron–positron plasmas \cite{PMID:25903920}. The development of the Extreme Light Infrastructure and proposals for the eXawatt Center for Extreme Light Studies \cite{2021QuEle,Khazanov2023XCELS} reflect ongoing efforts to enhance experimental capabilities for strong-field QED research.

Motivated by these experimental advances, vacuum pair production has been investigated using a broad range of analytical and numerical methods. Semiclassical approaches, including the worldline-instanton, Wentzel–Kramers–Brillouin, and imaginary-time methods, provide a physically intuitive framework for understanding nonperturbative pair production and allow the leading exponential behavior of pair-production probabilities to be evaluated through classical trajectories and complex actions \cite{PhysRevD.72.105004,Ritus,2005PAN,PhysRevLett.101.130404,PhysRevLett.104.250402,PhysRevD.83.065028}. Complementing these semiclassical treatments, quantum kinetic and quantum Vlasov approaches describe the real-time evolution of particle production, revealing the importance of memory effects and quantum coherence in pair creation \cite{PhysRevD.58.125015,PhysRevD.59.094005,PhysRevLett.102.150404}. Strong-field QED scattering-amplitude methods have also been applied to electron–positron pair production, with exact plane-wave Dirac solutions showing that pulse phase, separation, polarity, and ordering control azimuthal anisotropy and momentum-selective coherent enhancement in photon-induced pair production \cite{PhysRevD.98.036022,PhysRevD.101.016006}. Quantum-field evolution methods, including direct Dirac simulations, Klein–Gordon and Bogoliubov–Riccati approaches, and the Dirac–Heisenberg–Wigner (DHW) formalism, provide another framework for investigating pair production. Direct Dirac simulations of oscillating Sauter potentials reveal phase-sensitive pair yields, maximized by in-phase width and depth oscillations at low frequencies \cite{PhysRevA.100.022127}. Klein–Gordon and Bogoliubov–Riccati approaches reveal statistics-dependent Ramsey and multiple-slit interference, with selected momentum peaks scaling quadratically with pulse number \cite{PhysRevD.89.067701,PhysRevLett.108.030401}. Finally, DHW studies have shown that spatial localization, field oscillations, asymmetry, frequency chirping, and pulse shaping can suppress or enhance pair production and reshape momentum spectra \cite{PhysRevLett.107.180403,PhysRevD.97.036026,PhysRevD.100.016003,PhysRevD.101.096009,PhysRevD.104.036015,myr3-ql9h,Chen2024AsymmetricPulse,BAKE2024107486}.

Among these approaches, the covariant Wigner-function framework provides a phase-space formulation of quantum transport \cite{Vasak:1987um}, from which the DHW transport equations for spacetime-dependent fields have been derived \cite{PhysRevD.82.105026}. In spatially homogeneous fields, the DHW formalism has been related to quantum kinetic theory, with their equivalence further established for time-dependent electric fields of arbitrary polarization and multiple components \cite{PhysRevD.82.105026,PhysRevResearch.6.043009,LiJiang2026Nonadiabatic}. The DHW formalism is particularly suitable for spatially and temporally inhomogeneous external fields, enabling a real-time phase-space description of pair production and subsequent particle acceleration. DHW studies have shown that spatial confinement induces particle self-bunching and suppresses pair production in sufficiently narrow fields \cite{PhysRevLett.107.180403}, while ponderomotive forces split multiphoton momentum peaks \cite{PhysRevD.97.036026}. Reduced spatial extent also lowers total yields and weakens conventional temporal interference, while increasing the relative importance of dynamical assistance \cite{PhysRevD.100.016003}. Spatial asymmetry can enhance pair production and reshape momentum spectra \cite{myr3-ql9h}, whereas spatially oscillating fields can shift multiphoton peaks by transferring momentum to the produced particles, an effect not captured by locally homogeneous approximations \cite{PhysRevD.101.096009}. DHW studies of more complex electromagnetic field configurations have also revealed multiphoton interference structures in rotating electric fields \cite{PhysRevD.89.085001} and changes in pair yields and momentum distributions induced by inhomogeneous magnetic fields \cite{2016PhLB,PhysRevD.101.096003}. 

The relative ease of modifying temporal field profiles within the same DHW numerical framework, compared with the more involved treatment of spatial profiles through nonlocal integral operators, has facilitated more extensive investigations of temporal-field effects on pair production. For individual pulses, frequency chirps, carrier phases, temporal asymmetry, and envelope shape can significantly modify the momentum spectrum, interference pattern, and total particle yield \cite{mamutchrip,PhysRevD.104.016009,PhysRevD.104.036015,Mohamedsedik,Chen2024AsymmetricPulse,Xu2026SuperGaussian,OSMAN2023107195}. Oscillating Sauter fields and Gaussian or super-Gaussian pulse trains can further generate characteristic multiphoton peaks and coherent accumulation effects \cite{PhysRevA.100.022127}. In more complex configurations, the dynamically assisted Schwinger mechanism can significantly enhance pair production, while rapidly switched, bipolar, and multipulse fields can modify particle yields and momentum distributions through their temporal structures and pulse delays \cite{PhysRevD.100.016003,Chen2024AsymmetricPulse}. 
 
While temporal field structures have been widely studied, the spatial widths of individual components in multicomponent fields remain less systematically explored. Beyond the DHW framework, quantum-field simulations show that spatial separation weakens dynamical assistance between static and oscillating electric fields \cite{Ren2023SpatiallySeparated}, while locally constant field approximation calculations provide accurate pair-yield estimates for finite counterpropagating laser pulses using effective volume factors \cite{Tkachev2025VolumeFactors}. Worldline-instanton calculations reveal moiré interference in selected spacetime-separated pulse configurations \cite{DegliEsposti2025Spacetime}, whereas quantum-kinetic studies show that relative phases and temporal offsets strongly modify momentum spectra in superposed oscillating pulses \cite{Brass2026RelativePhase}. Within the DHW framework, previous studies have shown that spatial localization can induce particle self-bunching, while spatial focusing and oscillations reshape momentum spectra \cite{PhysRevLett.107.180403,PhysRevD.97.036026,PhysRevD.101.096009}. Finite spatial extent also modifies pulse interference \cite{PhysRevD.100.016003}, whereas spatial asymmetry influences particle yields and momentum distributions \cite{myr3-ql9h,kn4q-mv94}. Nevertheless, these studies do not systematically address a composite field formed by two co-centered Gaussian components with independently adjustable spatial-width parameters. To the best of our knowledge, this configuration has not been systematically explored within the $1+1$-dimensional DHW framework.

\begin{figure*}[t]
    \centering
    \includegraphics[width=\textwidth]{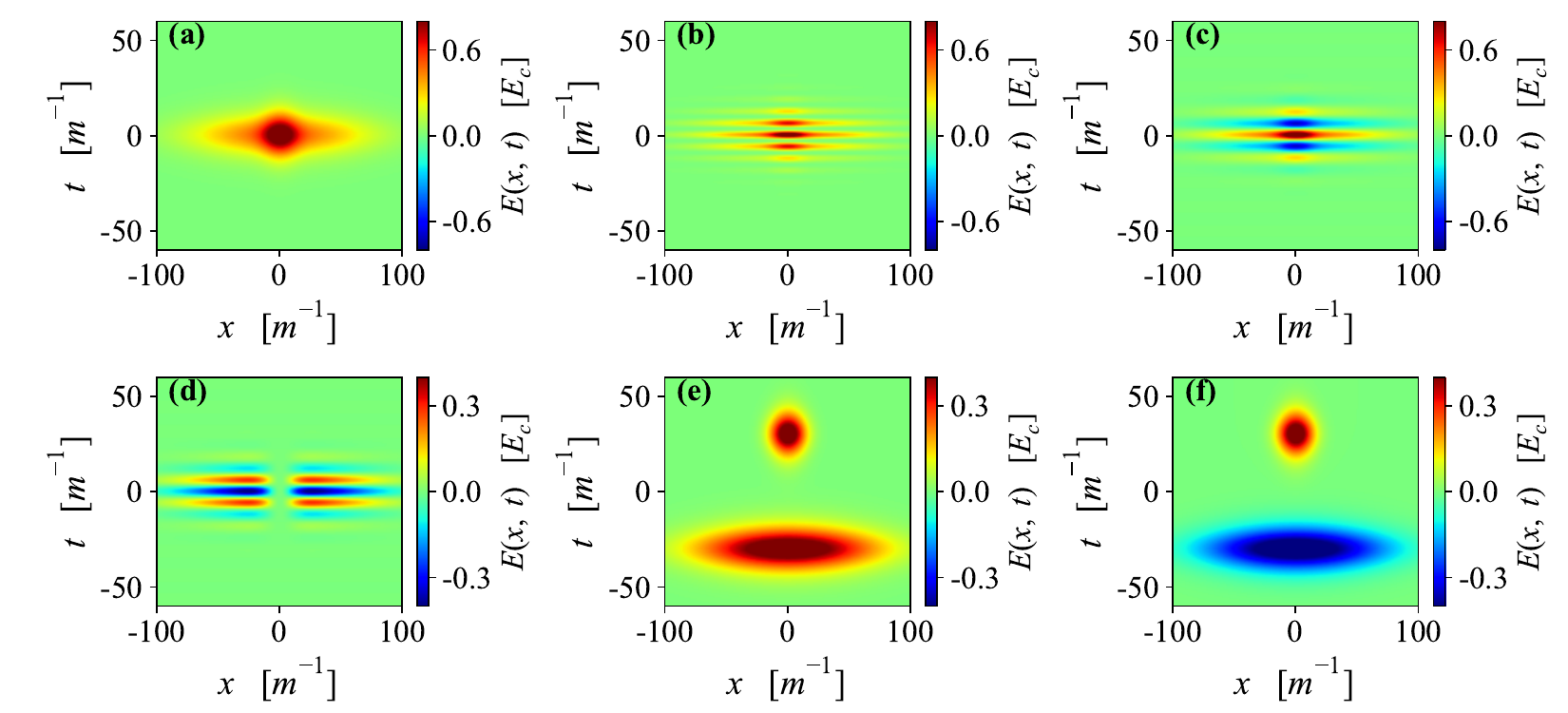}
    \caption{Composite electric-field profiles for (a) two simple pulses, (b) a truncated composite pulse, (c) two in-phase oscillating pulses, (d) two out-of-phase oscillating pulses, (e) two temporally symmetric pulses, and (f) two temporally antisymmetric pulses. The Gaussian spatial-width parameters are $\lambda_1=50\lambda_c$ and $\lambda_2=10\lambda_c$, and the pulse duration is $\tau=10\,m^{-1}$.}
    \label{laserfig}
\end{figure*}

In this work, we investigate Schwinger pair production in composite electric fields formed by two Gaussian components with independently adjustable spatial widths, $\lambda_1$ and $\lambda_2$. As illustrated in Fig.~\ref{laserfig}, we consider temporally overlapping simple, truncated, and oscillating composite pulses, as well as temporally separated symmetric and antisymmetric pulse pairs. For overlapping pulses, we examine how the difference between the spatial widths affects the momentum spectrum and phase-space particle distribution. For separated pulses, we investigate how the width of the delayed pulse influences the acceleration of particles produced by the earlier pulse. 

The remainder of this paper is organized as follows. Sec.~\ref{method} presents the external-field model, the DHW formalism, and the numerical strategy. Sec.~\ref{results} reports the numerical results for temporally overlapping pulses in Secs.~\ref{twos}--\ref{twoosci} and for temporally separated symmetric and antisymmetric pulses in Secs.~\ref{twosym} and~\ref{twoassy}, respectively. The main conclusions are summarized in Sec.~\ref{summary}.

\section{Pair production in $1+1$ dimensions}\label{method}
\subsection{External fields}\label{fields}

We investigate $e^-e^+$ pair production in $1+1$ dimensions using a two-pulse electric-field configuration with distinct spatial extensions. The field model is
\begin{equation}
\label{FieldModel}
E(x,t)=\epsilon_0 E_{\mathrm{cr}} e^{-\frac{x^{2}}{2\lambda_1^{2}}}g_1(t)
+\epsilon_0 E_{\mathrm{cr}} e^{-\frac{x^{2}}{2\lambda_2^{2}}}g_2(t),
\end{equation}
where $E_{\mathrm{cr}}$ is the critical field strength. The parameters $\lambda_1$ and $\lambda_2$ are the Gaussian spatial-width parameters of the two constituent pulses. Their independent variation is the defining feature of the present field model. When $\lambda_1=\lambda_2$, Eq.~\eqref{FieldModel} reduces to the commonly used combined-field configuration with a shared spatial envelope. For $\lambda_1\ne\lambda_2$, the two components sample different spatial regions and their superposition is generally non-Gaussian. 

\subsection{DHW formalism}
\label{DHWformalism}

To investigate vacuum pair production in the electric fields described by Eq.~\eqref{FieldModel}, we employ the real-time DHW formalism \cite{Vasak:1987um,PhysRevLett.107.180403}. This formalism has proved effective for analyzing pair production \cite{PhysRevLett.107.180403,2016PhLB,PhysRevD.89.085001}, particularly in spatially inhomogeneous external fields \cite{PhysRevLett.107.180403,2016PhLB}. This subsection briefly summarizes the formulation used in the present study. As throughout this article, the formulation is presented in natural units ($\hbar=c=1$), with all quantities expressed in terms of the electron mass $m$ and the electron Compton wavelength $\lambda_c$.

The DHW formalism involves the Dirac spinor fields $\Psi$ and $\bar{\Psi}$ and the vector potential $\mathbf{A}(x,t)$, which is treated as a classical external field. The spinor fields satisfy the Dirac equations \cite{PhysRevD.107.054502,PhysRevD.100.016003}
\begin{eqnarray}
\label{Eq_Dirac}
(\mathrm{i}\gamma^\mu\partial_\mu-e\gamma^\mu A_\mu(r)-m)\Psi(r)&=&0,\\
\bar{\Psi}(r)\left(\mathrm{i}\gamma^\mu\overleftarrow{\partial}_\mu+e\gamma^\mu A_\mu(r)+m\right)&=&0.
\end{eqnarray}

Within the DHW formalism, the Dirac equations can be transformed into transport equations. In $1+1$ dimensions, for $\mathbf{A}(\mathbf{x},t)=A(x,t)\mathbf{e}_x$, the relevant transport equations are \cite{mamutchrip,PhysRevD.100.016003,2016PhLB,PhysRevD.101.096003}
\begin{align}
&D_t\mathbbm{s}-2p_x\mathbbm{t}=0, \label{pde:1}\\
&D_t\mathbbm{v}_0+\partial_x\mathbbm{v}_1=0, \label{pde:2}\\
&D_t\mathbbm{v}_1+\partial_x\mathbbm{v}_0=-2m\mathbbm{t}, \label{pde:3}\\
&D_t\mathbbm{t}+2p_x\mathbbm{s}=2m\mathbbm{v}_1, \label{pde:4}
\end{align}
with the pseudodifferential operator
\begin{equation}
\label{pseudoDiff}
D_t=\partial_t+e\int_{-1/2}^{1/2}d\xi\,
E_x\!\left(x+\mathrm{i}\xi\partial_{p_x},t\right)\partial_{p_x}.
\end{equation}
Here, $\mathbbm{s}$, $\mathbbm{v}_0$, $\mathbbm{v}_1$, and $\mathbbm{t}$ are the Wigner components. The vacuum initial conditions are
\begin{equation}
\label{vacuum-initial}
\mathbbm{s}_{\mathrm{vac}}=-\frac{2m}{\omega},
\qquad
\mathbbm{v}_{1\,\mathrm{vac}}=-\frac{2p_x}{\omega},
\end{equation}
where $\omega(p_x)=\sqrt{p_x^2+m^2}$ is the single-particle energy. Vacuum-subtracted Wigner components are defined by
\begin{equation}
\mathbbm{w}_k^{v}(x,p_x,t)=\mathbbm{w}_k(x,p_x,t)-\mathbbm{w}_{k,\mathrm{vac}}(p_x).
\end{equation}

Following the Wigner-component normalization convention used in Refs.~\cite{PhysRevD.97.036026,PhysRevD.100.016003}, the momentum distribution is calculated as
\begin{equation}
\label{MS}
n(p_x,t)=\int\frac{dx}{2\pi}\,
\frac{\mathbbm{s}^{v}(x,p_x,t)+p_x\mathbbm{v}_1^{v}(x,p_x,t)}{\omega(p_x)},
\end{equation}
where
\begin{equation}
\frac{\mathbbm{s}^{v}(x,p_x,t)+p_x\mathbbm{v}_1^{v}(x,p_x,t)}{\omega(p_x)}
\end{equation}
is the phase-space particle-number density. The total particle yield is then
\begin{equation}
\label{totalYield}
N(t)=\int dp_x\,n(p_x,t).
\end{equation}

The spatial reduction used below depends on the field configuration. For a field with a single Gaussian spatial-width parameter $\lambda$, the reduced momentum spectrum and total yield are $\bar n=n/\lambda$ and $\bar N=N/\lambda$, respectively. For synchronized components with $\lambda_1\ne\lambda_2$, the composite spatial profile is fitted by a Gaussian of the form used in Eq.~\eqref{FieldModel}. The resulting width is denoted by $\lambda_{\mathrm{fit}}$ and is used to normalize the momentum spectrum and particle yield. For temporally separated pulses whose particle populations remain distinguishable, each contribution is normalized by the Gaussian width of its associated pulse. When the contributions overlap and cannot be assigned unambiguously to either pulse, no pulse-width normalization is applied. The same convention is used for the phase-space distributions shown below.

\subsection{Numerical strategy}
\label{strategy}

To solve Eqs.~\eqref{pde:1}--\eqref{pde:4}, the operator $D_t$ is decomposed into the ordinary time derivative $\partial_t$ and the field-dependent pseudodifferential part $\Delta$, such that $D_t=\partial_t+\Delta$. After subtracting the vacuum components, the evolution equations become
\begin{align}
\partial_t\mathbbm{s}^{v}&=-\Delta\mathbbm{s}^{v}-\Delta\mathbbm{s}_{\mathrm{vac}}+2p_x\mathbbm{t},\\
\partial_t\mathbbm{v}_0&=-\Delta\mathbbm{v}_0-\partial_x\mathbbm{v}_1^{v},\\
\partial_t\mathbbm{v}_1^{v}&=-\Delta\mathbbm{v}_1^{v}-\Delta\mathbbm{v}_{1\,\mathrm{vac}}-\partial_x\mathbbm{v}_0-2m\mathbbm{t},\\
\partial_t\mathbbm{t}&=-\Delta\mathbbm{t}-2p_x\mathbbm{s}^{v}+2m\mathbbm{v}_1^{v}.
\end{align}
The vacuum terms in the last equation cancel because $p_x\mathbbm{s}_{\mathrm{vac}}=m\mathbbm{v}_{1\,\mathrm{vac}}$.

The pseudodifferential contribution is evaluated by Fourier and inverse Fourier transforms. Let $w_{p_x}$ denote the Fourier variable conjugate to $p_x$ and let $\widehat{\mathbbm{w}}_k=\mathcal{F}_{p_x}[\mathbbm{w}_k]$. Then
\begin{equation}
\label{inifu}
\begin{aligned}
\Delta\mathbbm{w}_k(x,p_x,t)
={}&\mathcal{F}^{-1}_{p_x}\!\left[
\mathrm{i}e\,w_{p_x}\int_{-1/2}^{1/2}d\xi\,\right.\\
&\left.\quad\times E\!\left(x-\xi w_{p_x},t\right)
\widehat{\mathbbm{w}}_k(x,w_{p_x},t)
\right].
\end{aligned}
\end{equation}

When the two electric-field pulses have identical spatial profiles, a single integral is sufficient. For pulses with different spatial profiles, the corresponding contributions to Eq.~\eqref{inifu} must be evaluated separately. The spatial reflection symmetry of the external field can be exploited to simplify the numerical calculation. Since $E(x,t)=E(-x,t)$, the Wigner components have definite parity under $x\rightarrow -x$:
\begin{equation}
\begin{aligned}
\mathbbm{s}(x,p_x,t)&=\mathbbm{s}(-x,p_x,t),\\
\mathbbm{v}_0(x,p_x,t)&=-\mathbbm{v}_0(-x,p_x,t),\\
\mathbbm{v}_1(x,p_x,t)&=\mathbbm{v}_1(-x,p_x,t),\\
\mathbbm{t}(x,p_x,t)&=\mathbbm{t}(-x,p_x,t).
\end{aligned}
\end{equation}
Accordingly, $\mathbbm{s}$, $\mathbbm{v}_1$, and $\mathbbm{t}$ are even in $x$, while $\mathbbm{v}_0$ is odd, allowing the numerical domain to be restricted to $x\geq0$.

\section{Numerical results}
\label{results}

This section presents the numerical results for $e^-e^+$ pair production in the field configurations defined by Eq.~\eqref{FieldModel}. In each set of calculations, one of the spatial widths, $\lambda_1$ or $\lambda_2$, is held fixed while the other is varied. Three classes of temporal profiles are considered: temporally identical, temporally symmetric, and temporally antisymmetric pulses.

\subsection{Two temporally overlapping pulses}
\subsubsection{Two simple Gaussian pulses}
\label{twos}

\begin{figure}[t]
    \centering
    \includegraphics[scale=0.5]{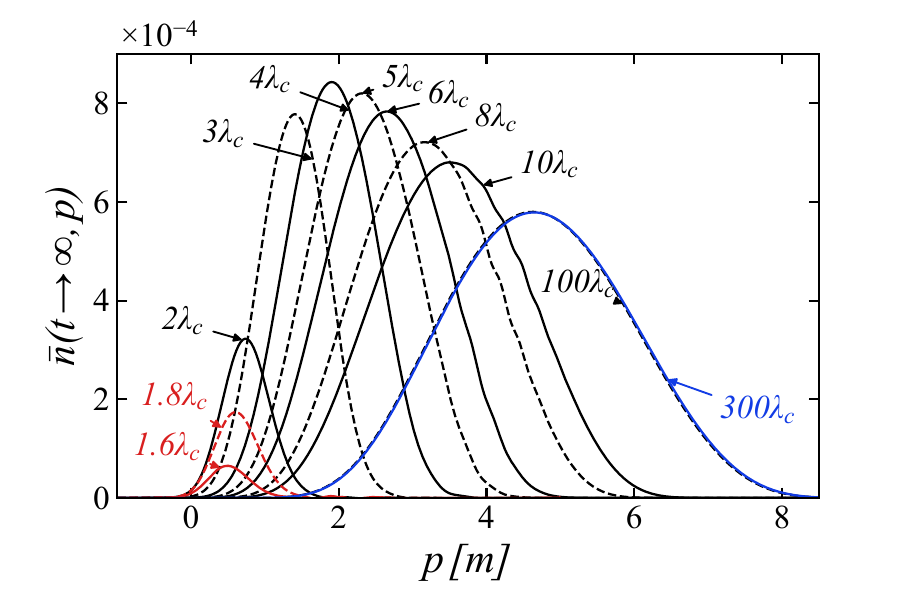}
    \caption{Reduced momentum spectra for two identical pulses described by Eq.~\eqref{FieldMode0}, for different values of the common spatial width $\lambda$. The parameters are $\epsilon_0=0.25$, $\lambda_1=\lambda_2=\lambda$, and $\tau=10\,m^{-1}$. The results agree with those of Ref.~\cite{PhysRevLett.107.180403}.}
    \label{self-bunching}
\end{figure}

Previous studies have examined the Schwinger effect for external fields with different spatial extensions and have demonstrated their influence on the momentum spectrum \cite{PhysRevD.100.016003}. Studying pair production in a wider range of spatial configurations can further improve our understanding of this effect. An advantage of the present field model is that the spatial extensions of the two pulses can be varied independently, allowing a variety of composite spatial profiles to be constructed.

We first consider two pulses with the same temporal profile,
$g_1(t)=g_2(t)=\operatorname{sech}^{2}(t/\tau)$. The external electric field is then
\begin{eqnarray}
\label{FieldMode0}
E(x,t)&=&\epsilon_0E_{\mathrm{cr}}e^{-\frac{x^2}{2\lambda_1^2}}
\operatorname{sech}^{2}\!\left(\frac{t}{\tau}\right)
+\epsilon_0E_{\mathrm{cr}}e^{-\frac{x^2}{2\lambda_2^2}}
\operatorname{sech}^{2}\!\left(\frac{t}{\tau}\right).
\end{eqnarray}

Before applying the method to fields with different spatial profiles, we calculate two identical Sauter pulses to validate the numerical implementation. We set $\epsilon_0=0.25$ and $\lambda_1=\lambda_2=\lambda$, with $\lambda=1.6\lambda_c$, $2\lambda_c$, $4\lambda_c$, $5\lambda_c$, $6\lambda_c$, $8\lambda_c$, $10\lambda_c$, $20\lambda_c$, $50\lambda_c$, $100\lambda_c$, and $300\lambda_c$. The reduced observables $\bar n(p,t)$ and $\bar N(t)$ are obtained by normalizing the momentum spectrum in Eq.~\eqref{MS} and the total yield in Eq.~\eqref{totalYield} by $\lambda$. Throughout this paper, the pulse duration is fixed at $\tau=10\,m^{-1}$. Because only the longitudinal momentum $p_x$ is nonzero, it is denoted by $p$ below. The reproduction of the momentum spectrum reported in Ref.~\cite{PhysRevLett.107.180403} validates the numerical implementation. The resulting reduced momentum spectra are shown in Fig.~\ref{self-bunching}.

\begin{figure}[t]
    \centering
    \includegraphics[width=0.5\textwidth]{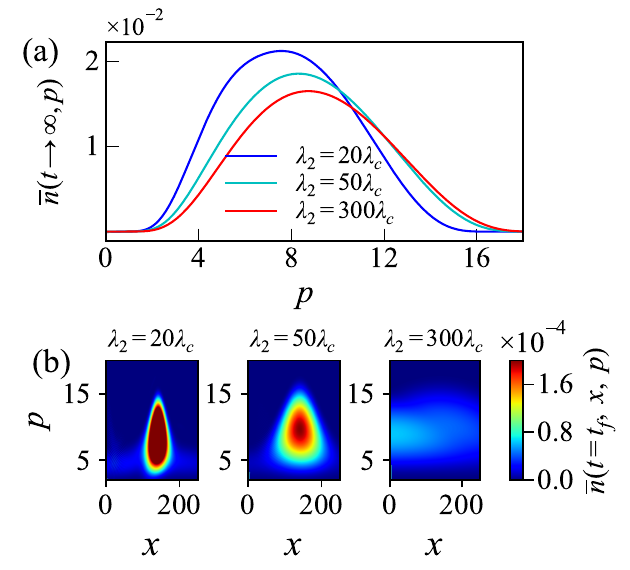}
    \caption{(a) Reduced momentum spectra and (b) reduced phase-space particle-density distributions for the field in Eq.~\eqref{FieldMode0}, with $\lambda_1=300\lambda_c$, $\lambda_2=20\lambda_c$, $50\lambda_c$, and $300\lambda_c$, and $\tau=10\,m^{-1}$. The distributions are normalized by the fitted Gaussian width $\lambda_{\mathrm{fit}}$.}
    \label{combined1}
\end{figure}

We next fix $\lambda_1=300\lambda_c$ and choose $\lambda_2=20\lambda_c$, $50\lambda_c$, and $300\lambda_c$ to examine the effect of the second spatial width. The corresponding reduced quantities are obtained using $\lambda_{\mathrm{fit}}$. Fig.~\ref{combined1}(a) shows a modest increase in the reduced particle number density as $\lambda_2$ decreases. At the same time, the spectrum shifts toward lower momenta and its shape changes.
To obtain further insight into the pair-production dynamics, we also examine the reduced phase-space particle density, as shown in Fig.~\ref{combined1}(b). The spatial extent of the particle distribution increases with $\lambda_2$, whereas the accessible momentum region becomes narrower. During pair production, particles are generated mainly in the strong-field region with small initial momenta and are subsequently accelerated by the electric field. In the spatially overlapping two-pulse configuration, the extent of this strong-field region is controlled by $\lambda_2$. Consequently, the spatial region occupied by the particles broadens as $\lambda_2$ increases. Particles with large momenta are mainly generated near the spatial origin at earlier times. Particles generated farther from the origin and moving outward leave the interaction region rapidly. By contrast, particles initially moving toward the origin from distant positions require more time to reach the strong-field region. Because the electric field changes during this interval, they experience a field weaker than its peak value. Consequently, reducing $\lambda_2$ increases the low-momentum particle density and broadens the momentum distribution.

\subsubsection{Cosine-squared-modulated Sauter-pulse combinations with truncated temporal structures}
\label{twotru}

\begin{figure}[t]
\centering
\includegraphics[width=0.5\textwidth]{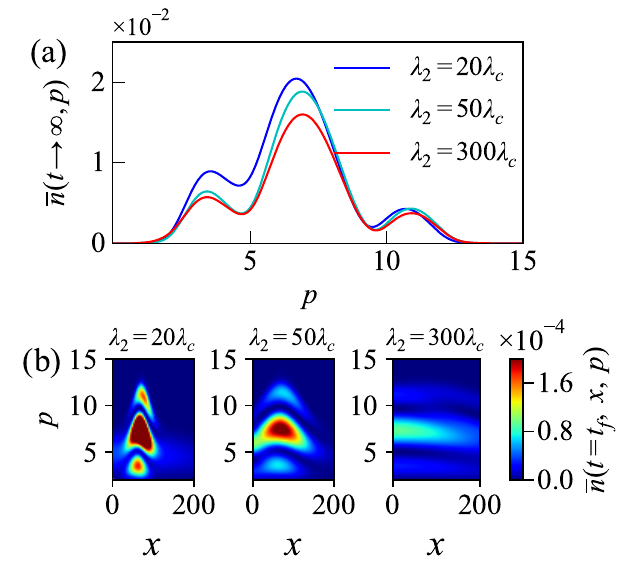}
\par\medskip
\includegraphics[width=0.5\textwidth]{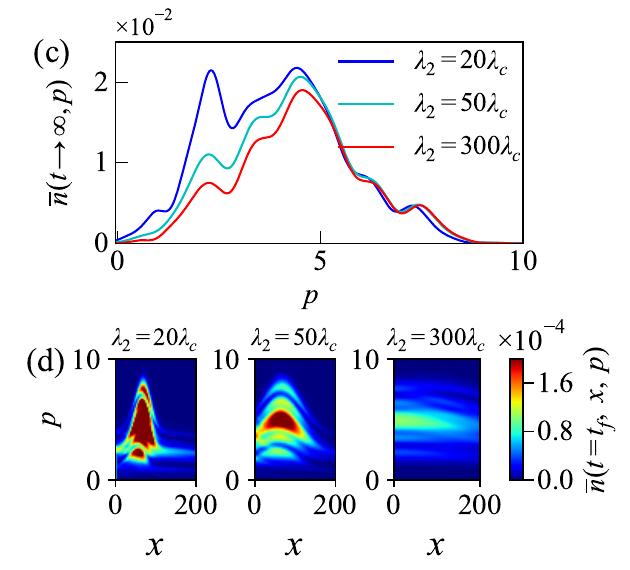}
\caption{(a) Reduced momentum spectra and (b) reduced phase-space particle-density distributions for an unmodulated Sauter component, $g_1(t)=\operatorname{sech}^2(t/\tau)$, combined with a cosine-squared-modulated component, $g_2(t)=\operatorname{sech}^2(t/\tau)\cos^2(\omega t)$. (c) Reduced momentum spectra and (d) reduced phase-space particle-density distributions when both components are cosine-squared modulated, $g_1(t)=g_2(t)=\operatorname{sech}^2(t/\tau)\cos^2(\omega t)$. In all panels, $\lambda_1=300\lambda_c$, $\lambda_2=20\lambda_c$, $50\lambda_c$, and $300\lambda_c$, $\tau=10\,m^{-1}$, and $\omega=0.5\,m$.}
\label{twotrun}
\end{figure}

We next consider cosine-squared-modulated Sauter-pulse combinations that generate truncated temporal structures in the composite field and examine the corresponding changes in the reduced momentum spectrum. The Gaussian spatial-width parameters are kept the same as in the preceding subsection: $\lambda_1=300\lambda_c$, while $\lambda_2=20\lambda_c$, $50\lambda_c$, and $300\lambda_c$. Figs.~\ref{twotrun}(a) and~\ref{twotrun}(b) show the results for $g_1(t)=\operatorname{sech}^2(t/\tau)$ and $g_2(t)=\operatorname{sech}^2(t/\tau)\cos^2(\omega t)$, whereas Figs.~\ref{twotrun}(c) and~\ref{twotrun}(d) show the results for $g_1(t)=g_2(t)=\operatorname{sech}^2(t/\tau)\cos^2(\omega t)$. In both configurations, $\omega=0.5\,m$. In these cases, the single peak in the reduced momentum spectrum splits into several peaks, and the entire distribution shifts toward lower momenta. The reduced particle density at low momenta increases as $\lambda_2$ decreases. The phase-space distributions contain structures that overlap in position but remain separated in momentum.

The particles are generated in the strong-field region and are accelerated during different temporal segments of the resulting truncated pulse. The different acceleration histories produce distinct final momenta. When an unmodulated Sauter component is combined with a cosine-squared-modulated component, the strong-field interval of the truncated pulse is divided into three effective temporal segments, producing a threefold splitting of the principal spectral peak. When both components are cosine-squared modulated, the resulting truncated pulse contains more temporal segments and therefore produces a larger number of peaks. Particles generated away from the spatial origin enter the strong-field region at different times and experience different weaker-field segments of the truncated pulse. As a result, the low-momentum peak becomes larger as $\lambda_2$ decreases.

\subsubsection{Two oscillating pulses}
\label{twoosci}

\begin{figure}[!tbp]
\centering
\includegraphics[width=0.5\textwidth]{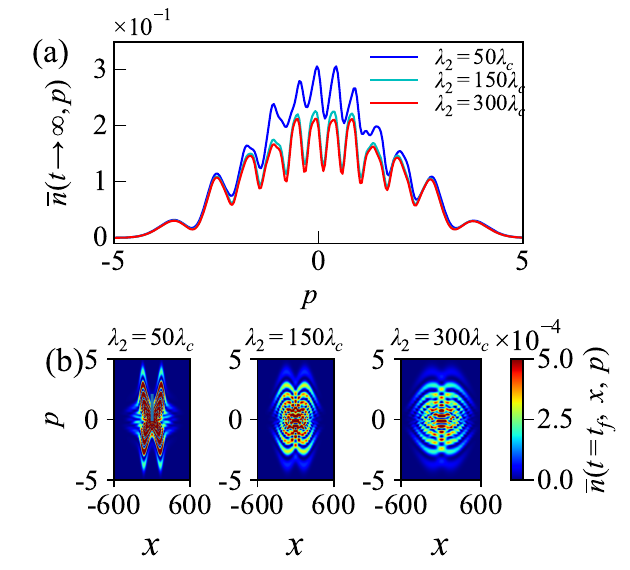}
\par\medskip
\includegraphics[width=0.5\textwidth]{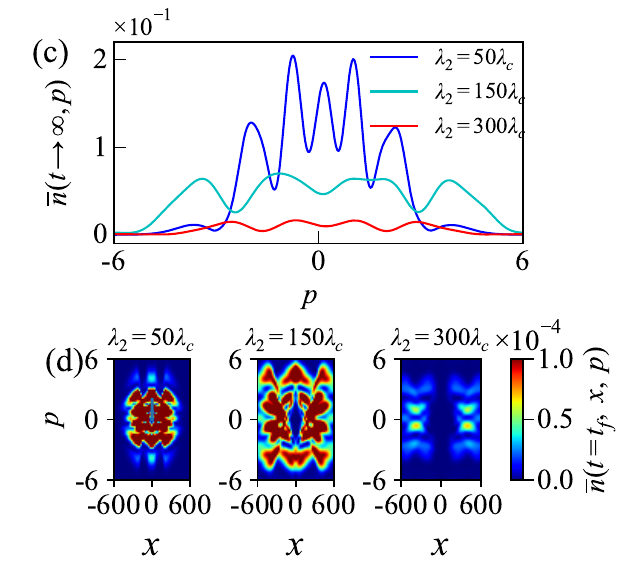}
\caption{(a) Reduced momentum spectra and (b) reduced phase-space particle-density distributions for in-phase oscillating components, $g_1(t)=g_2(t)=\operatorname{sech}^2(t/\tau)\cos(\omega t)$. (c) Reduced momentum spectra and (d) reduced phase-space particle-density distributions for out-of-phase components, $g_1(t)=-g_2(t)=\operatorname{sech}^2(t/\tau)\cos(\omega t)$. In all panels, $\lambda_1=300\lambda_c$, $\tau=10\,m^{-1}$, and $\omega=0.5\,m$, while $\lambda_2$ is varied.}
\label{simoci}
\end{figure}

In the preceding cases, the temporal profiles remain nonnegative, so particles of a given charge are predominantly accelerated in one direction. We now introduce oscillating fields to investigate how reversals of the electric-field direction affect pair production in different spatial configurations. Specifically, we take $g_1(t)=g_2(t)=\operatorname{sech}^2(t/\tau)\cos(\omega t)$, with $\omega=0.5\,m$. Figs.~\ref{simoci}(a) and~\ref{simoci}(b) show the reduced momentum spectra and reduced phase-space particle-density distributions for different values of $\lambda_2$ at fixed $\lambda_1=300\lambda_c$. Compared with the preceding nonoscillating configurations, the oscillating field exhibits an enhanced pair-production rate, and the reduced momentum spectra develop multiple peaks. Meanwhile, the phase-space distributions show particle accumulation around $x=0$ and $p=0$. Furthermore, the reduced particle density near $p=0$ decreases as $\lambda_2$ increases, whereas the high-momentum part changes only weakly.

The oscillating field can be approximately regarded as a sequence of alternating positive and negative temporal segments resembling truncated pulses. Particles created during different temporal segments experience different subsequent acceleration histories. In particular, reversals of the electric-field direction cause particles of a given charge to be accelerated in opposite momentum directions. Consequently, particles generated during different oscillation cycles can acquire different final momenta, contributing to the multiple peaks in the momentum spectrum. The repeated acceleration and deceleration may also contribute to oscillatory particle motion and particle accumulation around $x=0$ and $p=0$. Furthermore, the temporal oscillations can modify the nonadiabatic pair-production dynamics \cite{PhysRevLett.102.150404,PhysRevD.101.096009,PhysRevLett.101.130404}, which may contribute to the enhanced particle production in the oscillating configuration. The different responses of the high- and low-momentum regions to variations in $\lambda_2$ may be understood in terms of their different production regions. High-momentum particles mainly produced near $x=0$ during the early stages of the process, where the electric-field amplitude is independent of $\lambda_2$. This may explain why the high-momentum part of the spectrum remains relatively unchanged as $\lambda_2$ varies. In contrast, lower-momentum particles may be predominantly generated later in the weaker-field regions, where the spatial field profile is more sensitive to $\lambda_2$. This accounts for the more pronounced dependence of the reduced particle density near $p=0$ on $\lambda_2$.

We also consider the out-of-phase configuration
$g_1(t)=\operatorname{sech}^2(t/\tau)\cos(\omega t)$ and
$g_2(t)=\operatorname{sech}^2(t/\tau)\cos(\omega t+\pi)$.
The corresponding momentum spectra and phase-space distributions are shown in Figs.~\ref{simoci}(c) and~\ref{simoci}(d).
The $\pi$ phase difference reverses the sign of the second field component, leading to an exact cancellation at the spatial origin, where the two Gaussian envelopes have equal amplitudes. Away from the origin, however, the different spatial widths result in incomplete cancellation and generate two symmetric off-center residual-field regions. Pair production is therefore shifted toward these regions, while the alternating temporal field segments accelerate particles in opposite momentum directions, producing a multipeak spectrum. As $\lambda_2$ approaches $\lambda_1$, the two spatial profiles become increasingly similar, the residual field is weakened over a broader spatial range, and both the reduced momentum spectrum and the number of resolved peaks decrease. The phase-space distribution remains symmetric about $x=0$ and $p=0$, reflecting the spatial symmetry of the composite field and the alternating acceleration induced by the two out-of-phase components.

\subsection{Two temporally separated pulses}
\subsubsection{Two temporally symmetric pulses}
\label{twosym}

We next consider Schwinger pair production in two temporally symmetric pulses,
\begin{equation}
\label{sym}
\begin{split}
E(x,t)=&\,\epsilon_0E_{\mathrm{cr}}e^{-\frac{x^2}{2\lambda_1^2}}
\operatorname{sech}^2\!\left(\frac{t}{\tau}-\Delta_t\right)\\
&+\epsilon_0E_{\mathrm{cr}}e^{-\frac{x^2}{2\lambda_2^2}}
\operatorname{sech}^2\!\left(\frac{t}{\tau}+\Delta_t\right),
\end{split}
\end{equation}
where $\epsilon_0=0.5$ and $\Delta_t=3$.

For two spatially identical pulses, the momentum spectrum exhibits two regimes depending on the common spatial width. For $\lambda<20\lambda_c$, the reduced momentum spectrum has a single dominant peak. Above this range, an additional peak gradually appears at higher momentum \cite{PhysRevD.100.016003}. For sufficiently narrow pulses, particles produced by the earlier pulse can leave the strong-field region before the delayed pulse arrives, so the two particle populations may have similar final momenta and form a single peak. A broader delayed pulse, by contrast, can accelerate particles produced by the earlier pulse and separate the corresponding spectral contributions.

\begin{figure}[b]
    \centering
    \includegraphics[width=0.5\textwidth]{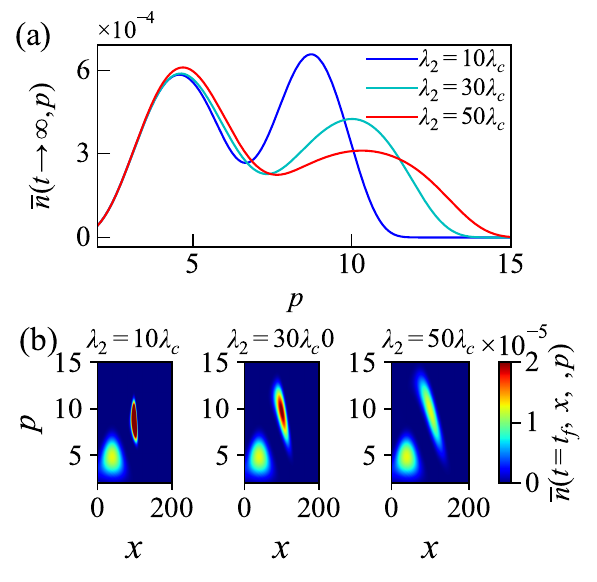}
    \caption{(a) Reduced momentum spectra and (b) reduced phase-space particle-density distributions for the pulse ordering in Eq.~\eqref{sym}, for different values of $\lambda_2$. The parameters are $\lambda_1=50\lambda_c$, $\epsilon_0=0.5$, $\tau=10\,m^{-1}$, and $\Delta_t=3$. Each distinguishable pulse-associated contribution is normalized by the spatial width of its associated pulse.}
    \label{l150l2235}
\end{figure}

We consider $\lambda_1=50\lambda_c$ and $300\lambda_c$ in turn while varying $\lambda_2$. We first discuss the case $\lambda_1=50\lambda_c$. Figs.~\ref{l150l2235}(a) and~\ref{l150l2235}(b) show, respectively, the reduced momentum spectra and reduced phase-space particle-density distributions for different values of $\lambda_2$. Two distinguishable contributions are visible in the momentum spectra. As $\lambda_2$ increases, the peak near $p\simeq13$ becomes broader and its maximum decreases, whereas the other peak changes only weakly. The phase-space distributions show that the two particle populations occupy different regions of phase space and remain distinguishable.

This behavior can be understood from the temporal ordering and spatial widths of the two pulses. The earlier pulse, centered at $t=-\Delta_t\tau$ and characterized by the width $\lambda_2$, first creates a particle population that is subsequently accelerated by the delayed pulse. The delayed pulse, centered at $t=\Delta_t\tau$ and characterized by the fixed width $\lambda_1$, also creates a second particle population. Since the two pulse-associated contributions remain distinguishable in momentum space, they are normalized separately: the contribution associated with the earlier pulse is divided by $\lambda_2$, whereas that associated with the delayed pulse is divided by $\lambda_1$. As $\lambda_2$ increases, the earlier pulse creates particles over a broader spatial region. When the delayed pulse arrives, particles closer to the spatial origin experience a stronger electric field and acquire larger momentum shifts, whereas particles farther from the origin undergo weaker acceleration. The resulting range of momentum shifts broadens the corresponding spectral contribution. Together with the normalization by $\lambda_2$, this broadening leads to a reduction in the maximum of the peak near $p\simeq13$. In contrast, the other peak is mainly associated with particles created directly by the delayed pulse. Because the spatial width and the other parameters of the delayed pulse remain fixed, this peak depends only weakly on $\lambda_2$.

\begin{figure}[t]
\centering
\includegraphics[width=0.5\textwidth]{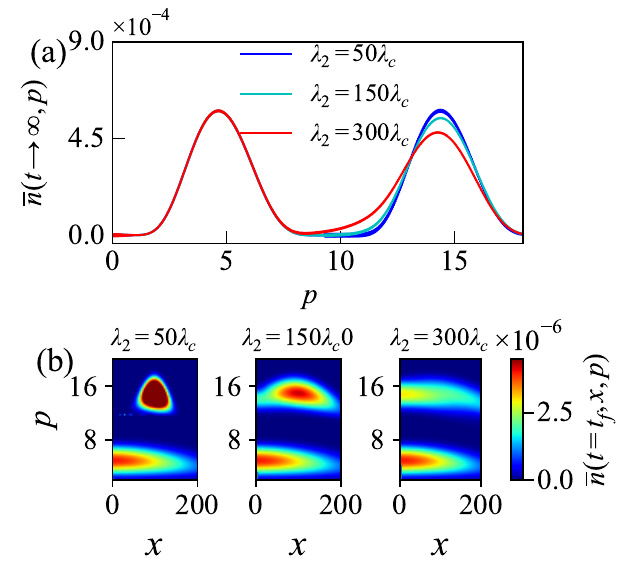}
\par\medskip
\includegraphics[width=0.5\textwidth]{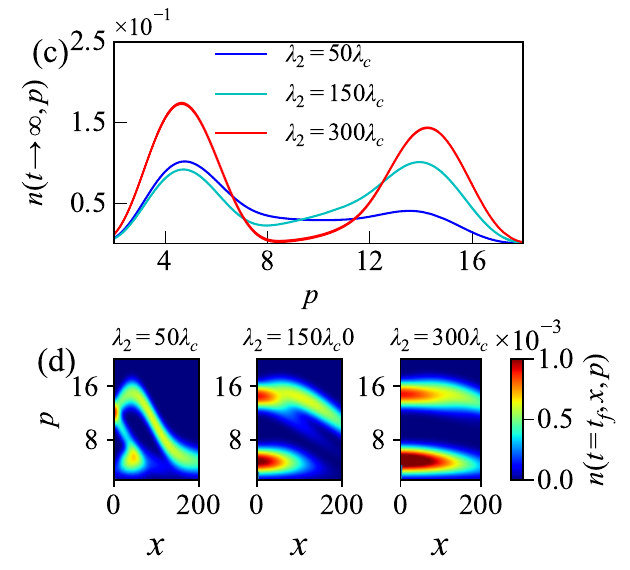}
\caption{(a) Reduced momentum spectra and (b) reduced phase-space particle-density distributions for $g_1(t)=\operatorname{sech}^2(t/\tau-3)$ and $g_2(t)=\operatorname{sech}^2(t/\tau+3)$. (c) Unnormalized momentum spectra and (d) unnormalized phase-space particle-density distributions for the reversed pulse order, $g_1(t)=\operatorname{sech}^2(t/\tau+3)$ and $g_2(t)=\operatorname{sech}^2(t/\tau-3)$. In all panels, $\lambda_1=300\lambda_c$, $\epsilon_0=0.5$, and $\tau=10\,m^{-1}$, while $\lambda_2$ is varied.}
\label{symmtric300}
\end{figure}

For $\lambda_1=300\lambda_c$, Figs.~\ref{symmtric300}(a) and~\ref{symmtric300}(b) show the reduced momentum spectra and reduced phase-space particle-density distributions, respectively, for different values of $\lambda_2$. Two distinct peaks appear near $p\simeq5$ and $p\simeq15$. As $\lambda_2$ varies, the positions and heights of the two peaks in the reduced spectra change only weakly, although clear differences remain in the corresponding phase-space distributions. The phase-space results indicate that the peak near $p\simeq15$ is mainly associated with the earlier pulse of width $\lambda_2$, whereas the peak near $p\simeq5$ is mainly associated with the delayed pulse of width $\lambda_1$.
This weak dependence of the reduced spectra on $\lambda_2$ can be understood from the relatively large spatial widths considered here. For $\lambda\gtrsim50\lambda_c$, pair production in the central interaction region approaches the behavior expected for a spatially homogeneous electric field \cite{PhysRevLett.107.180403}. 
Consequently, after the two distinguishable pulse-associated contributions are normalized by their respective spatial widths, their reduced peak heights and positions remain approximately unchanged. Nevertheless, the different values of $\lambda_2$ still modify the spatial extent of the earlier pulse and therefore produce visible differences in the phase-space distributions.

Reversing the temporal order of the two pulses gives
\begin{equation}
\label{symt}
\begin{split}
E(x,t)=&\,\epsilon_0E_{\mathrm{cr}}
e^{-\frac{x^2}{2\lambda_1^2}}
\operatorname{sech}^2\!\left(\frac{t}{\tau}+\Delta_t\right)\\
&+\epsilon_0E_{\mathrm{cr}}
e^{-\frac{x^2}{2\lambda_2^2}}
\operatorname{sech}^2\!\left(\frac{t}{\tau}-\Delta_t\right).
\end{split}
\end{equation}
Figs.~\ref{symmtric300}(c) and~\ref{symmtric300}(d) show the corresponding unnormalized momentum spectra and phase-space particle-density distributions. In this ordering, both the lower- and higher-momentum particle populations increase as $\lambda_2$ increases. The phase-space distributions also show that the contributions associated with the two pulses overlap in momentum space, particularly for $\lambda_2=50\lambda_c$. Because the particles in the overlapping region cannot be assigned unambiguously to either pulse, no pulse-resolved spatial-width normalization is applied to these results.

This behavior can be understood from the reversed production and acceleration sequence. The earlier, broad pulse of width $\lambda_1$ first creates particles over an extended spatial region. The delayed pulse of width $\lambda_2$ then produces an additional particle population while also accelerating particles created by the earlier pulse. For small $\lambda_2$, the delayed pulse strongly affects only particles located near the spatial origin, whereas particles farther from the origin experience weaker acceleration and remain at lower momenta. Their final momenta can therefore overlap with those of particles created directly by the delayed pulse. As $\lambda_2$ increases, the delayed field extends over a broader spatial region, accelerates a larger fraction of the preexisting particles, and produces pairs over a wider region. These effects account for the increase in both the lower- and higher-momentum particle populations.

\subsubsection{Two temporally antisymmetric pulses}
\label{twoassy}

\begin{figure}[t]
\centering
\includegraphics[width=0.5\textwidth]{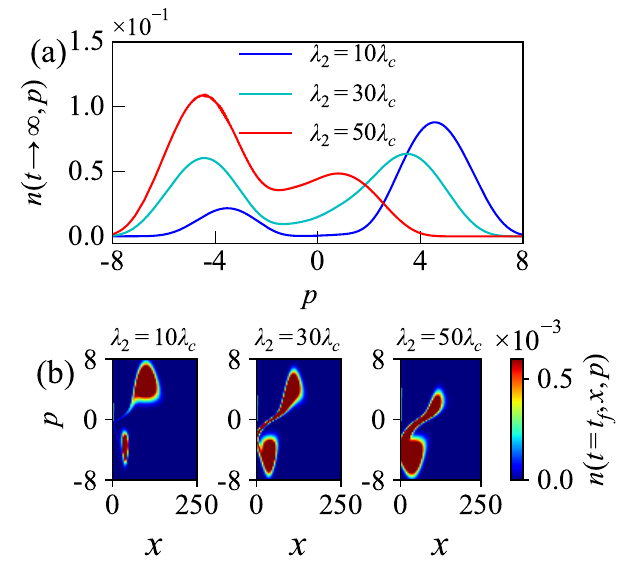}
\par\medskip
\includegraphics[width=0.5\textwidth]{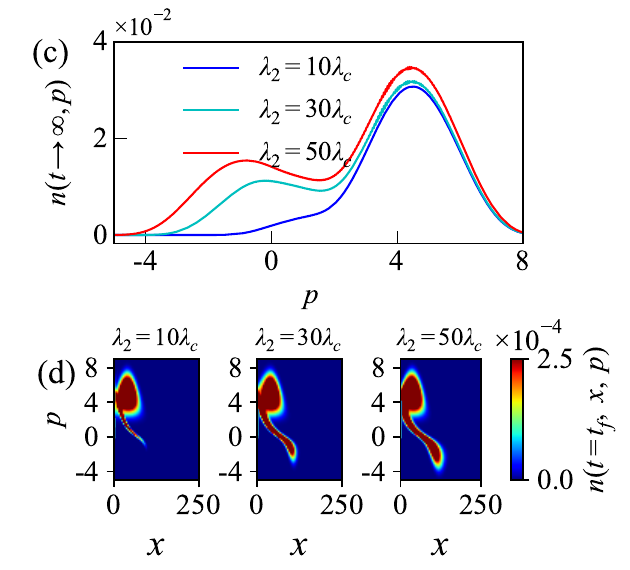}
\caption{(a) Momentum spectra and (b) phase-space particle-density distributions for the pulse ordering in Eq.~\eqref{asym}, in which the positive pulse of width $\lambda_1$ precedes the negative pulse of width $\lambda_2$. (c) Momentum spectra and (d) phase-space particle-density distributions for the reversed ordering, in which the negative pulse of width $\lambda_2$ precedes the positive pulse of width $\lambda_1$. In all panels, $\lambda_1=50\lambda_c$, $\lambda_2=10\lambda_c$, $30\lambda_c$, and $50\lambda_c$, $\epsilon_0=0.5$, $\tau=10\,m^{-1}$, and $\Delta_t=3$.}
\label{antisy}
\end{figure}

We next consider temporally antisymmetric pulse pairs, in which the two constituent pulses have opposite signs. The electric field is given by
\begin{equation}
\label{asym}
\begin{split}
E(x,t)=&\,\epsilon_0E_{\mathrm{cr}}
e^{-\frac{x^2}{2\lambda_1^2}}
\operatorname{sech}^2\!\left(\frac{t}{\tau}+\Delta_t\right)\\
&-\epsilon_0E_{\mathrm{cr}}
e^{-\frac{x^2}{2\lambda_2^2}}
\operatorname{sech}^2\!\left(\frac{t}{\tau}-\Delta_t\right).
\end{split}
\end{equation}
In this configuration, the positive pulse of width $\lambda_1$, centered at $t=-\Delta_t\tau$, precedes the negative pulse of width $\lambda_2$, centered at $t=\Delta_t\tau$. We fix $\lambda_1=50\lambda_c$ and consider $\lambda_2=10\lambda_c$, $30\lambda_c$, and $50\lambda_c$, with $\epsilon_0=0.5$ and $\Delta_t=3$.
Figs.~\ref{antisy}(a) and~\ref{antisy}(b) show the corresponding momentum spectra and phase-space particle-density distributions. As $\lambda_2$ increases, the negative-momentum peak becomes higher, whereas the positive-momentum peak decreases in height and shifts toward lower momenta. The phase-space distributions show that the particle populations associated with the two pulses overlap in momentum space. 

Since the parameters of the earlier positive pulse are fixed, it initially produces the same particle distribution for all values of $\lambda_2$. For small $\lambda_2$, the delayed pulse acts strongly only over a relatively narrow spatial region, so a substantial fraction of the particles retains positive momentum. As $\lambda_2$ increases, the delayed pulse extends over a broader spatial region and accelerates a larger fraction of the preexisting particles toward negative momentum. The delayed pulse also produces its own particle contribution, which overlaps with the redistributed particles in momentum space. These combined effects enhance the negative-momentum peak while reducing the positive-momentum contribution.

Figs.~\ref{antisy}(c) and~\ref{antisy}(d) show the momentum spectra and phase-space particle-density distributions for the reversed pulse ordering, in which the negative pulse of width $\lambda_2$ arrives before the positive pulse of width $\lambda_1$. For small $\lambda_2$, the final spectrum is dominated by the positive-momentum contribution, while the negative-momentum peak is weak. As $\lambda_2$ increases, the negative-momentum contribution becomes progressively more pronounced, and the spectrum develops a clearer two-peak structure. The corresponding phase-space distributions show that particles near the spatial origin are shifted toward positive momentum, whereas particles created farther from the origin are more likely to retain negative momentum. 

The earlier negative pulse first creates particles and accelerates them toward negative momentum. For small $\lambda_2$, this particle population is concentrated close to the spatial origin. When the delayed positive pulse of fixed width $\lambda_1$ arrives, most of these particles remain within its strong-field region and are efficiently accelerated toward positive momentum, producing a dominant positive-momentum peak. As $\lambda_2$ increases, the earlier pulse creates particles over a broader spatial region. Particles generated near the origin still experience strong acceleration from the delayed positive pulse and can be transferred to positive momentum. By contrast, particles generated farther from the origin experience a weaker delayed field and therefore retain part or all of their negative momentum, which enhances the negative-momentum peak. The delayed positive pulse also creates an additional particle population, and its contribution overlaps with that of the particles produced by the earlier pulse. The final spectrum therefore reflects both direct pair production by the two pulses and the momentum redistribution of the particles created by the first pulse.

\section{Conclusion}
\label{summary}

We have investigated Schwinger pair production in $1+1$ dimensions using the Dirac--Heisenberg--Wigner formalism for superposed electric fields composed of two Gaussian components with independently adjustable spatial widths, $\lambda_1$ and $\lambda_2$. Unlike configurations with a common spatial envelope, this model allows the effects of spatial-width disparity on pair production and particle dynamics to be examined. For temporally overlapping pulses, unequal spatial widths modify the composite field profile and the resulting momentum distributions. Decreasing $\lambda_2$ at fixed $\lambda_1$ generally enhances the low-momentum part of the reduced particle spectrum. Temporal modulation produces multipeak structures, while oscillating fields lead to a more pronounced accumulation of particles at low momenta. For out-of-phase oscillating components, cancellation at the spatial origin gives rise to off-center residual fields when the spatial widths differ, with the corresponding particle spectra becoming suppressed as $\lambda_2$ approaches $\lambda_1$.

For temporally separated pulses, the final momentum distribution depends sensitively on both the spatial widths and the temporal ordering of the constituent fields. The delayed pulse not only produces additional pairs but also accelerates particles created by the earlier pulse. Its spatial extent determines the fraction of these particles that undergo substantial subsequent acceleration, thereby modifying the spectral peak positions and relative strengths. For temporally symmetric pulses, varying the spatial widths changes the separation and broadening of the momentum peaks, whereas for antisymmetric pulses, the reversal of the field direction redistributes particles between positive- and negative-momentum regions. These results establish that the relative spatial widths of the constituent pulses provide an additional parameter for controlling pair production and momentum redistribution in multicomponent electric fields.

\begin{acknowledgments}
The author gratefully acknowledges Mamutjan Ababekri for insightful and valuable discussions. This work was supported by the Kashi University High-Level Talent Research Start-Up Fund (Grant No.~022024291) and the Kashi University Key Project (Grant No.~022024572).
\end{acknowledgments}

\bibliography{sample}
%\begin{thebibliography}{99}
%
%
%\end{thebibliography}
\end{document}